\documentclass[usenatbib,usegraphicx]{mn2e}
\usepackage{timesfonts}
\usepackage{journals}
\newcommand{\Y}{Y_\ell^m}

\newcommand{\bee}{\begin{eqnarray}}
\newcommand{\ene}{\end{eqnarray}}

\newcommand{\vx}{\mbox{\boldmath{$x$}}}

\newcommand{\di}{\mbox{\rm div}}

\newcommand{\vxi}{\mbox{\boldmath{$\xi$}}}

\begin{document}

\title[Asteroseismic Signatures of Activity Cycles]
{Asteroseismic Signatures of Stellar Magnetic Activity Cycles}

\author[Metcalfe et al.]{T.~S.~Metcalfe$^{1,2}$, W.~A.~Dziembowski$^{3,4}$, 
P.~G.~Judge$^1$, M.~Snow$^5$ \\
$^1$~High Altitude Observatory, National Center for Atmospheric Research,
Boulder, CO 80307-3000 USA      \\
$^2$~Scientific Computing Division, National Center for Atmospheric Research, 
Boulder, CO 80307-3000 USA  \\
$^3$~Warsaw University Observatory, Al.~Ujazdowskie 4, 00-478 Warsaw, Poland \\
$^4$~Copernicus Astronomical Centre, Bartycka 18, 00-716 Warsaw, Poland      \\
$^5$~Laboratory for Atmospheric and Space Physics, University of Colorado,
Boulder, Colorado 80309-0392 USA}

\maketitle

\begin{abstract} 
Observations of stellar activity cycles provide an opportunity to study 
magnetic dynamos under many different physical conditions. Space-based 
asteroseismology missions will soon yield useful constraints on the 
interior conditions that nurture such magnetic cycles, and will be 
sensitive enough to detect shifts in the oscillation frequencies due to 
the magnetic variations. We derive a method for predicting these shifts 
from changes in the Mg~{\sc ii} activity index by scaling from solar data. 
We demonstrate this technique on the solar-type subgiant $\beta$~Hyi, 
using archival {\it International Ultraviolet Explorer} spectra and two 
epochs of ground-based asteroseismic observations. We find qualitative 
evidence of the expected frequency shifts and predict the optimal timing 
for future asteroseismic observations of this star.
\end{abstract}



\section{Introduction}

Astronomers have been making telescopic observations of sunspots since the 
time of Galileo, gradually building an historical record showing a 
periodic rise and fall in the number of sunspots every $\sim$11 years. We 
now know that sunspots are regions with an enhanced local magnetic field, 
so this 11-year cycle actually traces a variation in surface magnetism. 
Attempts to understand this behavior theoretically often invoke a 
combination of differential rotation, convection, and meridional flow to 
modulate the field through a magnetic dynamo \citep[e.g., 
see][]{rem06,dg06}.

Although we can rarely observe spots on other solar-type stars directly, 
these areas of concentrated magnetic field produce strong emission in the 
Ca~{\sc ii}~H and K resonance lines in the optical, and the Mg~{\sc ii}~h 
and k lines in the ultraviolet. \cite{wil78} was the first to demonstrate 
that many solar-type stars exhibit long-term cyclic variations in their Ca 
{\sc ii} H~and~K emission, analogous to those seen in full-disc solar 
observations through the magnetic activity cycle. Early analysis of these 
data revealed an empirical correlation between the mean level of magnetic 
activity and the rotation period normalized by the convective timescale 
\citep{noy84a}, as well as a relation between the rotation rate and the 
period of the observed activity cycle \citep{noy84b}, which generally 
supports a dynamo interpretation.

Significant progress in dynamo modeling unfolded after helioseismology 
provided detailed constraints on the Sun's interior structure and 
dynamics. These observations also established that variations in the mean 
strength of the solar magnetic field lead to significant shifts 
($\sim$0.5~$\mu$Hz) in the frequencies of even the lowest-degree p-modes 
\citep{lw90,sal04}. These shifts can provide independent constraints on 
the physical mechanisms that drive the solar dynamo, through their 
influence on the outer boundary condition for the pulsation modes. They 
are thought to arise either from changes in the near-surface propagation 
speed due to a direct magnetic perturbation \citep{gol91}, or from a 
slight decrease in the radial component of the turbulent velocity in the 
outer layers and the associated changes in temperature \citep{dg04,dg05}.

Space-based asteroseismology missions, such as MOST \citep{wal03}, CoRoT 
\citep{bag06}, and Kepler \citep{jcd07} will soon allow additional tests 
of dynamo models using other solar-type stars \citep[see][]{cha07}. High 
precision time-series photometry from MOST has already revealed 
latitudinal differential rotation in two solar-type stars 
\citep{cro06,wal07}, and the long-term monitoring from future missions is 
expected to produce asteroseismic measurements of stellar convection zone 
depths \citep{mct00,vce06}. By combining such observations with the 
stellar magnetic activity cycles documented from long-term surveys of the 
Ca~{\sc ii} or Mg~{\sc ii} lines, we can extend the calibration of dynamo 
models from the solar case to dozens of independent sets of physical 
conditions.

The G2 subgiant $\beta$~Hyi is the only solar-type star that presently has 
both a known magnetic activity cycle \citep{dra93} and multiple epochs of 
asteroseismic observations \citep{bed01,bed07}. In this paper we reanalyze 
archival {\it International Ultraviolet Explorer} (IUE) spectra for an 
improved characterization of the magnetic cycle in this star, and we use 
it to predict the activity related shifts in the observed radial p-mode 
oscillations. We compare these predictions with recently published 
asteroseismic data, and we suggest the optimal timing of future 
observations to maximize the amplitude of the expected p-mode frequency 
shifts.


\section{Archival IUE spectra\label{MGSEC}}

The activity cycle of $\beta$~Hyi was studied by \cite{dra93}, who used 
high resolution IUE data of the Mg~{\sc ii} resonance lines over 11 years, 
from June 1978 to the end of October 1989. They considered these data to 
be consistent with a cycle period between 15 and 18 years. Since the work 
of \citeauthor{dra93}, a significant number of additional IUE spectra were 
obtained by E.~Guinan from early 1992 to the end of 1995. Our analysis of 
all of these spectra reveals the beginning of a new cycle in 1993-1994. 

In 1997, the IUE project reprocessed the entire database using improved 
and uniform reduction procedures (``NEWSIPS''). Using the NEWSIPS merged 
high resolution extracted spectra, we have reanalyzed the entire IUE 
dataset containing useful echelle data of the Mg~{\sc ii} lines. Data were 
excluded when the NEWSIPS software mis-registered the spectral orders, when 
continuum data near 279.67~nm were saturated, or when continuum data were 
more than $1\sigma$ below the mean (to reject additional poorly registered 
spectral orders) or more than $1.5\sigma$ above the mean.

The classic definition of the Mg~{\sc ii} index \citep{hs86} uses wing 
irradiances at 276 and 283~nm. The wings in their formulation had to be so 
far away from the cores due to the 1.1~nm spectral bandpass of their 
instrument. In the IUE spectra, pixels at those wavelengths are saturated, 
so the photospheric reference levels need to be measured much closer to 
the emission cores. \cite{sm05} have shown that at moderate resolution the 
variability of the inner wings of the Mg~{\sc ii} absorption feature is 
very similar to the variability of the classic wing irradiances. 
Therefore, we can construct a modified Mg~{\sc ii} index using only the 
unsaturated IUE data that still captures the full chromospheric 
variability. 

\begin{figure}
\centering
\includegraphics[width=\columnwidth]{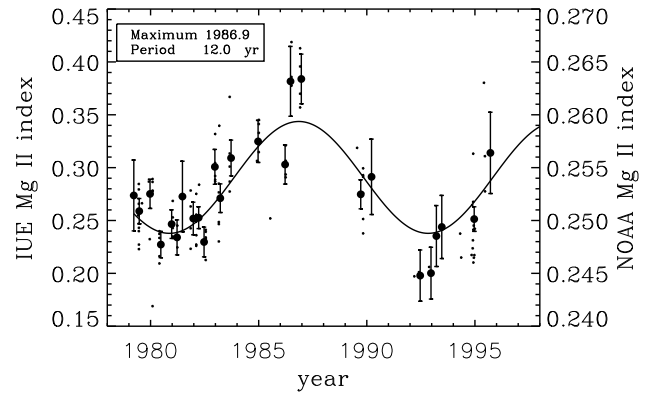}
\caption[fig1.eps]{ Core to wing ratios of the summed Mg~{\sc ii} h and k 
lines determined from IUE high dispersion observations of $\beta$~Hyi 
(small points) and 3-month seasonal averages with the uncertainties used 
in the evaluation of $\chi^2$ shown as error bars (large points). The 
curve is an optimized simple sinusoid fit obtained using a genetic 
algorithm applied to the seasonally averaged data, intended only to 
estimate the period and phase of the stellar activity cycle, which is 
listed in the legend. The IUE index is shown on the left, while the 
corresponding NOAA index is shown on the right.\label{fig1}}
\end{figure}

The chromospheric line cores (0.14 and 0.12~nm wide bandpasses centered at 
279.65 and 280.365~nm, in vacuo) and two bands in the photospheric wings 
of the lines (0.4~nm wide bands, edge-smoothed with cosine functions, 
centered at 279.20 and 280.70~nm) were integrated, and the ratio of total 
core to total wing fluxes was determined. Figure~\ref{fig1} shows the core 
to wing indices determined from each usable spectrum from 1978 to the end 
of 1995. The post-1992 data permit us to revise the cycle period estimate 
downwards to $\sim$12.0 years, with more confidence than was previously 
possible. This period was derived by fitting a simple sinusoid to the data 
using the genetic algorithm PIKAIA \citep{cha95}. The optimal fit yields 
minima at 1980.9 and 1992.8, and a maximum at 1986.9. The fit suggests 
that the next maximum occurred in 1998.8, a minimum in 2004.8, and a 
future maximum predicted for 2010.8. The reduced $\chi^2$ of the fit was 
calculated using flux uncertainties for individual IUE observations of 
7\%, estimated from the variation in the ratios of the two wing fluxes, 
which vary far less than this in the SOLSTICE solar data. This reduced 
$\chi^2$ has a minimum value of 0.77. The probability of such a value 
occurring at random is 24\%, whereas a $\chi^2$ of $1.1$ has a random 
probability of 76\%. The $\chi^2=1.1$ hypersurface contours suggests that 
the uncertainties are roughly $\pm1$~yr for the phase and 
$^{+3.0}_{-1.7}$~yr for the period, making our new period estimate 
marginally consistent with the range quoted by \cite{dra93}. The $\chi^2$ 
contours are ovals because these uncertainties are correlated, allowing us 
to set the following formal limits on the epochs of maximum: 
1986.1--1988.0, 1997.5--2002.4, and 2007.8--2017.8. These large 
uncertainties reinforce the need for an activity cycle monitoring program 
specifically for the southern hemisphere.

To compare the IUE Mg~{\sc ii} index measurements to the National Oceanic 
\& Atmospheric Administration (NOAA) composite data of solar 
activity\footnote{http://www.sec.noaa.gov/ftpdir/sbuv/NOAAMgII.dat}, we 
must determine the appropriate scaling factor. The SOLar-STellar 
Irradiance Comparison Experiment (SOLSTICE) on the SOlar Radiation and 
Climate Experiment \citep[SORCE;][]{mrw05} measures the solar irradiance 
every day, and has a resolution of 0.1~nm in this region. We convolved the 
IUE spectra with the SOLSTICE instrument function and then measured the 
wings and cores of both solar and stellar data in exactly the same way. In 
particular, we used 279.15-279.35~nm as the blue wing, and 
280.65-280.85~nm for the red wing. The emission cores were defined as 
279.47-279.65~nm and 280.21-280.35~nm. We determined the relation between 
the SOLSTICE modified Mg~{\sc ii} index and the NOAA long-term record 
using a standard linear regression method \citep[see][]{sno05,vie04}. 
Since the modified IUE data has the same bandpass as the SOLSTICE data, 
the scaling factors derived from SOLSTICE solar data will also apply to 
the stellar IUE data\footnote{To transform between indices: NOAA = 0.211 + 
0.0708 SOLSTICE; SOLSTICE = 0.297 + 1.11 IUE; NOAA = 0.232 + 0.079 IUE}. 
The Mg~{\sc ii} index for $\beta$~Hyi scaled to the NOAA composite data is 
shown on the right axis of Figure~\ref{fig1}. For the analysis in 
Section~\ref{SHIFTSEC}, we adopt a full amplitude of $\Delta i_{\rm 
MgII}=0.015$ in the NOAA index.


\section{Scaling p-mode shifts from solar data \label{SHIFTSEC}} 

In general, we can evaluate activity related frequency shifts from the 
variational expression,
\begin{equation}
\Delta\nu_j={\int d^3\vx{\cal K}_j{\cal S}\over 2I_j\nu_j},
\end{equation}
where 
\begin{equation}
I_j=\int d^3\vx\rho|\vxi|^3\equiv R^5\bar\rho\tilde I
\end{equation}
is the mode inertia, $j\equiv(n,\ell,m)$, and we need to know both the 
source ${\cal S}(\vx)$ and the corresponding kernels ${\cal K}(\vx)$. The 
source must include the direct influence of the growing mean magnetic 
field, as well as its indirect effect on the convective velocities and 
temperature distribution. Separate kernels for these effects were 
calculated by \cite{dg04}, but there is no theory available to calculate 
the combined source. Moreover, it is unclear whether the model of 
small-scale magnetic fields, adopted from \cite{gol91}, is adequate. 
Therefore, we will attempt to formulate an extrapolation of the solar 
p-mode frequency shifts based on changes in the Mg~{\sc ii} activity index 
measured for the Sun and for $\beta$~Hyi in Section~\ref{MGSEC}.

For p-modes, the dominant terms in all of the kernels are proportional to 
$|\di\vxi|^2$. Thus, we write
\begin{equation}
{\cal K}_j(\vx)=|\di\vxi_j|^2=q_j(D)\Y,
\end{equation}
where $D$ is the depth beneath the photosphere. A model-dependent 
coefficient will be absorbed into the source, which we write in the form
\begin{equation}
{\cal S}(\vx)=\sum_{k=0}{\cal S}_k(D)P_{2k}(\cos\theta).
\end{equation}
Solar data imply that ${\cal S}$ is strongly concentrated near the 
photosphere. Therefore data from all p-modes, regardless of their $\ell$ 
value, may be used to constrain ${\cal S}$. We might also expect that the 
source normalization is correlated with the Mg~{\sc ii} index.

If we want to calculate $\Delta\nu_{j}$ according to Eq.~(1), we need all 
terms of ${\cal S}$ up to $k=\ell$. To assess the solar ${\cal S}$, we 
have measurements of the centroid shifts and the even-$a$ coefficients 
\citep[see][]{dg04}. For the $\ell=0$ modes we only need to know the $k=0$ 
term, and for this the centroid data are sufficient. Let us begin with 
this simple case.

\subsection{Radial modes}

Theoretical arguments and the observed pattern of solar frequency changes 
suggest that the dominant source must be localized near the photosphere. 
Therefore, it seems reasonable to try to fit the measured p-mode frequency 
shifts by adopting
\begin{equation}
{\cal S}_0(D)=1.5\times10^{-11}A_0\delta(D-D_c)~\mu{\rm Hz}^2,
\end{equation}
with adjustable parameters $A_0$ and $D_c$. The numerical coefficient is 
arbitrary, and was chosen for future convenience. With Eqs.~(3-5), we get 
from Eq.~(1) 
\begin{equation}
\Delta\nu_j=A_0{R\over M}Q_j(D_c),
\end{equation}
where $R$ and $M$ (as well as $L$ below) are expressed in solar units, 
frequencies are expressed in $\mu$Hz, and
\begin{equation}
Q_j=1.5\times10^{-11}{q_j\over\nu_j\tilde I_j}.
\end{equation}

The solar values of $A_0$ and $D_c$ can be determined by fitting the 
centroid frequency shifts $\Delta\nu_j$ from SOHO MDI data for p-modes 
with various spherical degrees, $\ell$. Since at $n=1$ the approximation 
inherent in Eq.~(3) is questionable, we use data only for the higher 
orders.

For the Sun, we have 
\begin{equation}
A_{0,\odot}(D_c)=\overline{\left(w_j\Delta\nu_j\over Q_j\right)},
\end{equation}
where $\Delta\nu_j$ are the measured shifts and $w_j$ are the relative 
weights. The values of $Q_j$ are calculated from a solar model. The best 
value of $D_c$ is that which minimizes the dispersion,
\begin{equation}
\sigma(D_c)=\sqrt{\overline{\left[w_j\left(A_{0,\odot}-{\Delta\nu_j
\over Q_j}\right)^2\right]}}.
\end{equation}
It is also reasonable to assume that $A_0$ should be proportional to the 
change in the Mg~{\sc ii} activity index, $\Delta i_{\rm MgII}$, and 
that $D_c$ is proportional to the pressure scale height at the 
photosphere. Thus, we have
\begin{equation}
A_0=A_{0,\odot}{\Delta i_{\rm MgII}\over\Delta i_{{\rm MgII},\odot}}
\end{equation}
and 
\begin{equation}
D_c\propto H_p=D_{c,\odot}L^{0.25}{R^{1.5}\over M}.
\end{equation}

\begin{figure}
\centering
\includegraphics[width=\columnwidth]{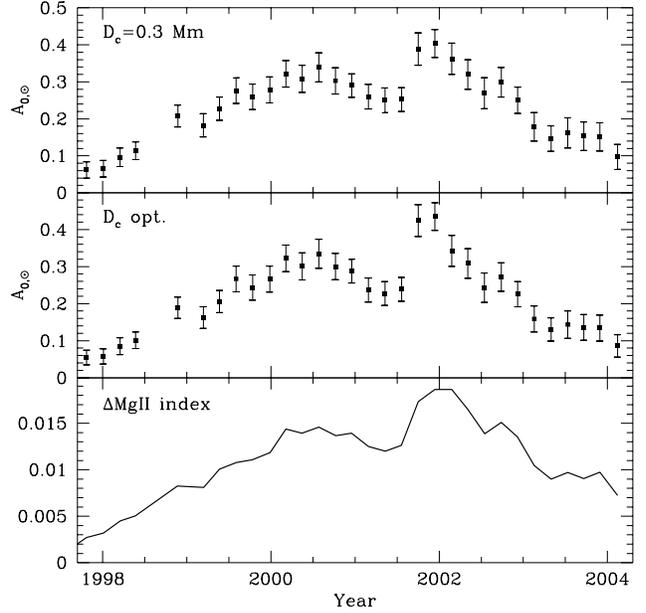}
\caption[fig2.eps]{Determinations of $A_{0,\odot}$ from SOHO MDI data 
with $D_c$ fixed at 0.3~Mm (top panel), and using the optimal value of 
$D_c$ for each set (middle panel), with the corresponding changes in the 
Mg~{\sc ii} index from the NOAA composite data (bottom panel).
\label{fig2}} 
\end{figure} 

To determine $A_{0,\odot}$, we used SOHO MDI frequencies for all p$_n$ 
modes with $n>1$, $\ell$ from 0 to $\sim$181, and $\nu$ between 2.5 and 
4.2 mHz. The data were combined into 38 sets, typically covering 
0.2~years. We averaged the frequencies from the first 5 sets, 
corresponding to solar minimum (1996.3-1997.3), and subtracted them from 
the frequencies in subsequent sets to evaluate $A_{0,\odot}$ using 
Eq.~(8). The results are shown in Figure~\ref{fig2}, where the points in 
the top panel were obtained at fixed $D_c=0.3$~Mm, which is representative 
of the highest activity period. In the middle panel, the value of $D_c$ 
was determined separately for each set and the error bars represent the 
dispersion. In the bottom panel, we show the corresponding changes in the 
solar Mg~{\sc ii} index calculated from NOAA composite data. A tight 
correlation between $A_{0,\odot}$ and $\Delta i_{{\rm MgII},\odot}$ is 
clearly visible.

\begin{figure}
\centering
\includegraphics[width=\columnwidth]{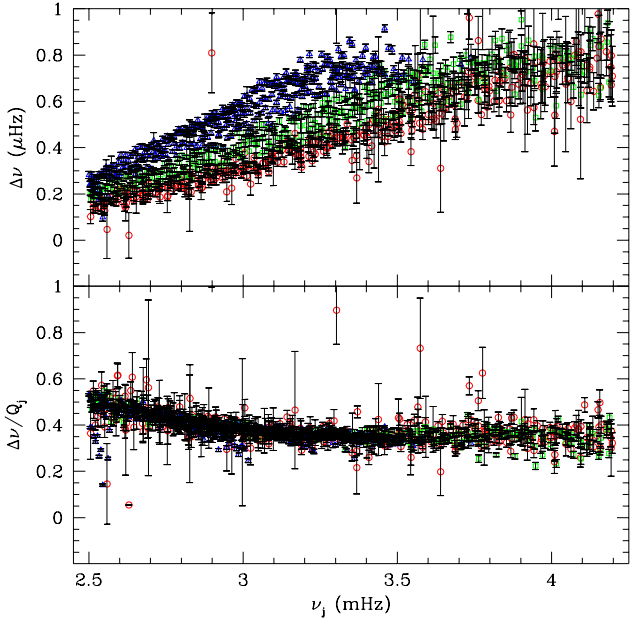}
\caption[fig3.eps]{The observed p-mode frequency shifts averaged from four 
sets of data obtained near the solar maximum in 2002.0 (top panel), and 
the same shifts normalized by $Q_j$ (bottom panel) showing that most of 
the frequency and $\ell$-dependence is included in our parametrization. 
Different symbols show the roughly equal number of modes with $\ell\le30$ 
(circles), $31\le\ell\le75$ (squares), and $76\le\ell\le181$ 
(triangles).\label{fig3}}
\end{figure}

Although the optimum value of $D_c$ is weakly correlated with the activity 
level, the dispersion changes very little between $D_c=0.2$ and 0.4~Mm, so 
we fixed the value of $D_{c,\odot}$ to 0.3~Mm. In Figure~\ref{fig3}, we 
show the quality of the fit to the observed frequency shifts for selected 
p-modes using Eq.~(6) with the adopted value of $D_c$. The upper panel 
shows the frequency shifts averaged from four sets of data near the 
activity maximum in 2002.0, while the lower panel shows $\Delta\nu_j/Q_j$. 
Note that most of the frequency and $\ell$-dependence appears to have been 
fit by our parametrization. The slight rise at frequencies below 3~mHz 
could be eliminated by allowing a spread of the kernel toward lower 
depths. However, since the signal is more significant at higher 
frequencies, we believe that adding a finite radial extent would be an 
unnecessary complication.

There were two activity maxima during solar cycle 23. The first was 
centered near 2000.6 and the second at 2002.0. The average values of 
($A_{0,\odot},\Delta i_{{\rm MgII},\odot}$) are (0.3116, 0.0135) for five 
data sets around the first maximum and (0.3669, 0.0178) for four data sets 
around the second maximum. For future applications, we adopt 
$A_{0,\odot}/\Delta i_{{\rm MgII},\odot}=22$.

With this specification, we get from Eqs.~(6) and (10) 
\begin{equation} 
\Delta\nu_j=3.3\times10^{-10}\Delta i_{\rm MgII} {R\over 
M}{q_j(D_c)\over\nu_j\tilde I_j}, 
\end{equation} 
where 
\begin{equation} 
D_c=0.3L^{0.25}{R^{1.5}\over M}\mbox{ mM} 
\end{equation}
and again frequencies are expressed in $\mu$Hz, while $R$, $M$, and $L$ 
are in solar units. This is our expression for predicting the radial 
p-mode frequency shifts on the basis of changes in the NOAA composite 
Mg~{\sc ii} index. In Table~\ref{tab1}, we list the frequency shifts 
($\Delta\nu_j$) calculated from Eq.~(12) for the radial modes of 
$\beta$~Hyi observed by \cite{bed07}, adopting $\Delta i_{\rm 
MgII}=0.015$. The mode parameters $q_j$ and $\tilde I_j$ were calculated 
from a model of $\beta$~Hyi generated using the Aarhus STellar Evolution 
Code \citep[ASTEC;][]{jcd82}.

\subsection{Non-radial modes}

Now from Eqs.~(1) and (4). we have
\begin{equation}
\Delta\nu_{nlm}={\sum_{k=0}^\ell{\cal S}_k\kappa_{k,lm}
\over 2I_{nl}\nu_{nl}},
\end{equation}
where,
\begin{equation}
\kappa_{k,lm}=\int d\theta d\phi|\Y|^2P_{2k}(\cos\theta)\sin\theta.
\end{equation}
As in Eq.~(5), we can adopt
\begin{equation}
{\cal S}_k(D)=1.5\times10^{-11}A_k\delta(D-D_{c,k})~\mu{\rm Hz}^2.
\end{equation}

For $k>0$, the solar amplitudes $A_k$ and effective depths, $D_{c,k}$ can 
be determined by fitting measurements of shifts in the $a_{2k}$ 
coefficients. The relation is
\begin{equation}
\Delta a_{2k,\ell m}=A_kZ_{k,\ell}Q_{k,n\ell}(D_{c,k}),
\end{equation}
where
\begin{equation}
Z_{k,\ell}=(-1)^k{(2k-1)!!\over k!}{(2\ell+1)!!\over(2\ell+2k+1)!!}
{(\ell-1)!\over(\ell-k)!}
\end{equation}
\citep[cf.][their Eq.~2]{dg04}, and
\begin{equation}
Q_{k,n\ell}=1.5\times10^{-11}{q_{n\ell}(D_{c,k})\over\nu_j
\tilde I_{n\ell}}
\end{equation}
(compare to our Eq.~7).

The prediction of frequency shifts for non-radial modes requires an 
additional assumption of the same scaling for all required $A_k$ 
amplitudes, which amounts to assuming the same {\it Butterfly diagram} as 
observed on the Sun. Moreover, since the shifts depend on $|m|$ and 
multiplets are not expected to be resolved, we need to adopt the 
inclination angle ($i$) to correctly weight the contributions from all of 
the components. Since we do not know $i$ for $\beta$~Hyi, we restrict our 
numerical predictions to the radial modes.

\begin{table}
\centering
\caption[tab1]{Predicted radial p-mode frequency shifts between activity 
maximum and minimum for $\beta$~Hyi, calculated with Eq.~(12) and adopting 
$\Delta i_{\rm MgII}=0.015$. Frequencies are from Table~1 of \cite{bed07}. 
\label{tab1}}
\begin{tabular}{@{}ccc@{}}
\hline
$n$& Frequency ($\mu$Hz) & $\Delta\nu_j$ ($\mu$Hz)\\
\hline
 13 &  833.72 & 0.061 \\
 14 &  889.87 & 0.091 \\
 15 &  945.64 & 0.116 \\
 16 & 1004.21 & 0.139 \\
 17 & 1062.06 & 0.168 \\
 18 & 1118.93 & 0.199 \\
 19 & 1176.48 & 0.234 \\
\hline
\end{tabular}
\end{table}


\section{Asteroseismic Observations}

The detection of solar-like oscillations in $\beta$~Hyi was first reported 
by \cite{bed01}, and later confirmed by \cite{car01}. These two detections 
of excess power were based on data obtained during a dual-site campaign 
organized in June 2000 using the 3.9-m Anglo-Australian Telescope (AAT) at 
Siding Spring Observatory and the 1.2-m Swiss telescope at the European 
Southern Observatory (ESO) in Chile. Both sets of observations measured a 
large frequency separation between 56-58~$\mu$Hz, but neither was 
sufficient for unambiguous identification of individual oscillation modes.

Nearly 30 individual modes in $\beta$~Hyi with $\ell$=0-2 were detected 
during a second dual-site campaign organized in September 2005, and 
reported by \cite{bed07}. The authors also reanalyzed the combined 2000 
observations using an improved extraction algorithm for the AAT data, 
allowing them to identify some of the same oscillation modes at this 
earlier epoch. Motivated by the first tentative detection of a systematic 
frequency offset between two asteroseismic data sets for $\alpha$~Cen~A 
\citep[0.6$\pm$0.3~$\mu$Hz;][]{fle06}, they compared the two epochs of 
observation for $\beta$~Hyi and found the 2005 frequencies to be 
systematically lower than those in 2000 by 0.1$\pm$0.4~$\mu$Hz, consistent 
with zero but also with the mean value in Table~\ref{tab1}.

A comparison of the individual modes from these two data sets (T.~Bedding, 
private communication) allows a further test of our predictions. Of the 14 
modes that were detected with S/N\,$>4$ in both 2000 and 2005, only one 
was known to be a radial ($\ell=0$) mode, while four had $\ell=1$, three 
had $\ell=2$, two were mixed modes, and four had no certain 
identification. Without a known inclination, we can only calculate the 
shifts for radial modes, but the magnitude of the shift is largest at high 
frequencies (see Table~\ref{tab1}). Fortunately, the radial mode that is 
common to both data sets ($\ell=0$, $n=18$) has a frequency above the peak 
in the envelope of power, improving our chances of measuring a shift. The 
best estimate of the mode frequency from each data set comes from the 
noise-optimized power spectrum, since this maximizes the S/N of the 
observed peaks. The noise-optimized frequency for the $\ell=0$, $n=18$ 
mode was 1119.06 and $1118.89~\mu$Hz in the 2000 and 2005 data sets, 
respectively. Considering the quoted uncertainty for this mode from 
Table~1 of \cite{bed07}, the frequency was 0.17$\pm$0.62~$\mu$Hz lower in 
2005 than in 2000, again consistent with zero but similar to the predicted 
shift for this mode in Table~\ref{tab1}.


\section{Discussion}

Our reanalysis of archival IUE spectra for $\beta$~Hyi allows us to test 
our predictions of the relationship between the stellar activity cycle and 
the systematic frequency shift measured from multi-epoch asteroseismic 
observations. The optimal period and phase of the activity cycle from 
Section~\ref{MGSEC} suggest that $\beta$~Hyi was near magnetic minimum 
(2004.8) during the 2005 observations (2005.7), while it was descending 
from magnetic maximum (1998.8) during the 2000 campaign (2000.5). The 
systematic frequency shift of 0.1$\pm$0.4~$\mu$Hz reported by \cite{bed07} 
between these two epochs, and the observed shift of 0.17$\pm$0.62~$\mu$Hz 
in the only radial mode ($\ell=0$, $n=18$) common to both data sets are 
not statistically significant. They are both nominally in the direction 
predicted by our analysis of the activity cycle (lower frequencies during 
magnetic minimum) and they have approximately the expected magnitude 
(cf.~Table~\ref{tab1}), but the formal uncertainties on the period and 
phase of the activity cycle do not permit a definitive test.


Future asteroseismic observations of $\beta$~Hyi would sample the largest 
possible frequency shift relative to the 2005 data if timed to coincide 
with the magnetic maximum predicted for $2010.8^{+7}_{-3}$. Long-term 
monitoring of the stellar activity cycles of this and other southern 
asteroseismic targets (e.g.~$\alpha$~Cen A/B, $\mu$~Ara, $\nu$~Ind), which 
are not included in the Mt.~Wilson sample, would allow further tests of 
our predictions. For asteroseismic targets that have known activity cycles 
from long-term Ca~{\sc ii} H and K measurements (e.g.~$\epsilon$~Eri, 
Procyon), it would be straightforward to calibrate our predictions to this 
index from comparable solar observations.

While our current analysis involves a simple scaling from solar data, 
future observations may allow us to refine magnetic dynamo models by 
looking for deviations from this scaling relation and attempting to 
rectify the discrepancies. By requiring the models to reproduce the 
observed activity cycle periods and amplitudes---along with the resulting 
p-mode shifts and their frequency dependence for a variety of solar-type 
stars at various stages in their evolution---we can gradually provide a 
broader context for our understanding of the dynamo operating in our own 
Sun.


\section*{ACKNOWLEDGMENTS}

We would like to thank D.~Salabert for inspiring this work with an HAO 
colloquium on low-degree solar p-mode shifts in May 2005, Keith MacGregor 
and Margarida Cunha for thoughtful discussions, and the Copernicus 
Astronomical Centre for fostering this collaboration during a sponsored 
visit in September 2006. We also thank the SOHO/MDI team, and especially 
Jesper Schou for easy access to the solar frequency data, Tim Bedding for 
providing frequency data for $\beta$~Hyi, and J{\o}rgen 
Christensen-Dalsgaard for the use of his stellar evolution code. This work 
was supported in part by an NSF Astronomy \& Astrophysics Fellowship under 
award AST-0401441, by Polish MNiI grant No.~1~P03D~021~28, and by NASA 
contract NAS5-97045 at the University of Colorado. The National Center for 
Atmospheric Research is a federally funded research and development center 
sponsored by the U.S. National Science Foundation.

\end{document}